\documentclass[conference]{IEEEtran}
\IEEEoverridecommandlockouts
\usepackage{cite}
\usepackage{amsmath,amssymb,amsfonts}
\usepackage{algorithmic}
\usepackage{multirow}
\usepackage{graphicx}
\usepackage{array}
\usepackage{enumitem}
\newcolumntype{L}[1]{>{\raggedright\arraybackslash}p{#1}}
\usepackage[caption=false, font=footnotesize]{subfig} 
\usepackage{booktabs}
\graphicspath{{Figures/}}

\usepackage{comment}
\usepackage{textcomp}
\usepackage{url}
\usepackage{xcolor}

\def\BibTeX{{\rm B\kern-.05em{\sc i\kern-.025em b}\kern-.08em
    T\kern-.1667em\lower.7ex\hbox{E}\kern-.125emX}}

\renewcommand{\baselinestretch}{0.97}

\begin{document}

\title{Link-Level Performance Comparison of 5G Beamforming: Reciprocity-Based SVD vs. Codebook-Based Precoding }

\author{
\IEEEauthorblockN{
S.M. Haider Ali Shuvo\IEEEauthorrefmark{1}\IEEEauthorrefmark{3},
Seda Dogan-Tusha\IEEEauthorrefmark{1},
and Monisha Ghosh\IEEEauthorrefmark{1}}
\vspace{3pt}
\IEEEauthorblockA{
\IEEEauthorrefmark{1}University of Notre Dame, USA,
\IEEEauthorrefmark{3}Corresponding Author Email: sshuvo@nd.edu }
}

\maketitle

\begin{abstract}

Downlink (DL) precoding in mid-band MIMO relies on uplink (UL) sounding reference signals (SRS) with reciprocity-based singular value decomposition (SVD) or channel state information reference signals (CSI-RS) with codebook-based precoding matrix indicator (PMI) feedback. Although SVD precoding offers unquantized beamforming, it depends on a channel estimate from SRS transmitted at the limited user equipment (UE) power, whereas Type-I PMI is quantized but derived from a high-power CSI-RS based channel estimate. In this paper, we compare SRS-SVD and CSIRS-PMI for single-user MIMO through an end-to-end, 3GPP-compliant link-level simulation at 3.75~GHz with 100~MHz bandwidth, using a 192 antenna element gNB mapped into 32 ports, the 3GPP Release~19 UE antenna model, and ray-traced (RT) channels in Manhattan and Chicago. We show that neither scheme is uniformly superior: CSIRS-PMI serves weak and shadowed users better and reduces outage, whereas SRS-SVD achieves higher spectral efficiency (SE) and enables more multi-layer transmission for users with strong channels. The results on the Clustered Delay Line (CDL)-C statistical channel also confirm this crossover. We further show that by utilizing a simple logistic regression based on the received power of the standard CSI reference signal (CSI-RSRP) and the channel quality indicator (CQI), the gNB can select the optimal transmission scheme with around 90\% accuracy, achieving an SE that closely approaches an ideal per-user selection. \looseness=-1

\end{abstract}

\vspace{-0.7em}

\section{Introduction} \label{sec:introduction}

Multiple-input multiple-output (MIMO) in the mid-band portion of FR1 (e.g., 3.3--4.2~GHz) is the main capacity layer of commercial 5G networks, where active antenna systems (AAS) with up to 64 transceivers (64T64R) and 192 antenna elements serve users through beamforming and spatial multiplexing~\cite{Nokia_ExtremeMIMO}. 
However, the achievable gain of large arrays depends not only on the array itself but also on the beamforming approach adopted at the base station (gNB), namely how downlink (DL) channel state information (CSI) is acquired and how the precoder is derived from it. \looseness=-1

3GPP NR specifies two fundamental mechanisms for DL CSI acquisition and precoding~\cite{3GPP_TS38214, 10375688}:

\textbf{i) CSI-RS with PMI Feedback (CSIRS-PMI)}: The gNB transmits channel state information reference signals (CSI-RS), and the receiver, i.e., user equipment (UE), estimates the DL channel and feeds back a precoding matrix indicator (PMI) selected from a standardized codebook, ranging from the low-resolution Type-I single-panel codebook to the high-resolution Type-II, enhanced Type-II, and port-selection codebooks introduced in Releases~15--19~\cite{3GPP_TS38214, Ning_PMI_Tutorial}.

\textbf{ii) SRS with SVD Precoding (SRS-SVD)}: In time-division duplex (TDD) systems, the UE transmits sounding reference signals (SRS), and the gNB estimates the uplink (UL) channel.
Leveraging channel reciprocity, the gNB derives the DL channel directly from this UL estimate after applying a calibration factor.
The gNB then performs singular value decomposition (SVD) of the estimated DL channel and uses the dominant right-singular vectors as the precoder, which, in contrast to PMI, is not restricted to a finite codebook~\cite{Dahlman_5GNR}.

Each approach has distinct limitations.
CSIRS-PMI is quantized: the reported precoder is restricted to a finite codebook, which cannot fully capture the spatial structure of the channel and therefore limits the beamforming gain. SRS-based SVD avoids quantization, but its accuracy depends on SRS transmitted at the limited UE power, which is split across UE antenna ports and spread over the full bandwidth. As a result, SRS reaches the gNB with substantially lower per-subcarrier power than CSI-RS reaches the UE. The two schemes therefore embody a fundamental trade-off between \emph{precoder resolution} and \emph{estimation quality}, and the better choice is expected to vary from user to user with link budget and the propagation environment. \looseness=-1 

Prior work has examined this trade-off from different perspectives. In~\cite{Samsung_MassiveMIMO_2020}, SRS-based and PMI-based precoding are compared through system-level simulations with statistical channels, reporting cell-average throughput gains, while the per-user performance of the two schemes remains to be quantified.
Other studies \cite{Lee_DLReconstruction_2021,Becirovic_Reciprocity_Feedback} combine SRS with quantized feedback to reconstruct the DL channel, but consider smaller arrays with normalized channel gains 
and thereby removing the UL--DL link-budget imbalance that limits SRS in practice.
To the best of our knowledge, a 3GPP-compliant, per-user comparison of SRS-SVD and CSIRS-PMI that jointly accounts for commercial-scale gNB AAS, realistic UE antenna model, and site-specific propagation remains largely unexplored, as does a practical rule for selecting between them using standard metrics available at the gNB. 
In this paper, we address this gap through an end-to-end, 3GPP-compliant link-level evaluation. The main contributions of this work are summarized below: \looseness=-1
\begin{itemize}[nosep, leftmargin=*]
    \item We develop a link-level simulation framework that combines the complete 5G NR PDSCH transmit and receive chains with ray-traced (RT) channels in two urban deployments, Manhattan and Chicago. The framework uses a realistic link budget, a 192-element gNB virtualized into 32 CSI-RS ports, and the 3GPP Release~19 UE antenna model.
    \item We show that neither SRS-SVD nor CSIRS-PMI is uniformly superior. CSIRS-PMI serves weak and shadowed users better and reduces outage, whereas SRS-SVD achieves higher spectral efficiency (SE) and higher transmission rank for users with strong channels.
    \item We propose a simple feature-based rule in which a logistic regression on standard CSI Reference Signal Received Power (CSI-RSRP) and CQI reports (or SRS-RSRP, available directly at the gNB without UE feedback) predicts the better scheme with about 90\% accuracy and closely matches the SE of an oracle that always picks the best scheme.
\end{itemize}

\section{Simulation Methodology}

\subsection{gNB and UE Configuration}
\begin{figure}[t]
  \centering
  \subfloat[\label{fig:gnb_ant}]{\includegraphics[height=1.7in]{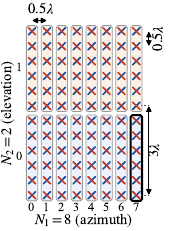}}
  \subfloat[\label{fig:ue_ant}]{\includegraphics[height=1.6in]{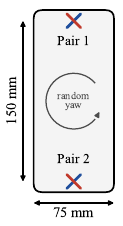}}
  \caption{Antenna models (blue: $+45^\circ$, red: $-45^\circ$ polarization): (a) gNB dual-polarized UPA with $192$ elements  
  where the bold box marks one subarray; 
  (b) 3GPP Release~19 UE model with two cross-polarized pairs.}
  \label{fig:antennas}
  \vspace{-1em}
\end{figure}

The gNB's AAS is modeled as a $\pm45^\circ$ dual-polarized uniform planar array (UPA) with 192 antenna elements~\cite{ITU-R_5D989_Annex4.4}. The array consists of $8\times12$ element positions in the azimuth and elevation dimensions, respectively, with a uniform spacing of $0.5\lambda$ in both dimensions. Consistent with the 3GPP Release~19 IMT parameters, the maximum directional gain of each element is set to 6.4~dBi, including ohmic losses~\cite{3GPP_TR38922}, while the element radiation pattern follows the model in~\cite{3GPP_TR38901}. In all simulations, the gNB is deployed at a height of 25~m with a $6^\circ$ downtilt, in line with the 3GPP Urban Macro (UMa) scenario~\cite{3GPP_TR38901, ITU-R_5D989_Annex4.4}. The physical array is logically partitioned via subarray virtualization, grouping subarray of $1\times6$ elements (one in azimuth and six in elevation) to one logical port~\cite{Ericsson_MobiFone_2024}. The six elements within each subarray are coherently combined with equal-magnitude weights. Consequently, the 192 physical antenna elements are mapped to 32 effective CSI-RS ports, corresponding to $N_1=8$ azimuth ports, $N_2=2$ elevation ports, and two polarization components. The effective port spacing is $0.5\lambda$ in azimuth and $3\lambda$ in elevation, as shown in Fig.~\ref{fig:gnb_ant}.

The 3GPP Release 19 introduces a new UE antenna model \cite{11389146} based on a standard handheld form factor of 150mm $\times$ 75mm. Although the model defines up to eight candidate antenna locations along the device edges, the active placement depends on the frequency band. For our mid-band simulation, we assume the antennas are separated by the full length of the device.
The UE is therefore modeled with two pairs of $\pm45^\circ$ cross-polarized elements separated by 150~mm, giving four UE ports as illustrated in Fig. \ref{fig:ue_ant}. Each element has a maximum gain of 2.3~dBi accounting for ohmic loss, a half-power beamwidth of $125^\circ$ in both azimuth and elevation, and a side-lobe attenuation of 22.5~dB, following the Release~19 specifications. Each UE is assigned a random yaw orientation to reflect the arbitrary ways users hold their devices, while the pitch is set to zero.

\subsection{Ray Tracing Channel}
\begin{figure}[t]
  \centering
  \subfloat[Manhattan simulation scenario \label{F7a}]{%
    \includegraphics[width=0.48\linewidth]{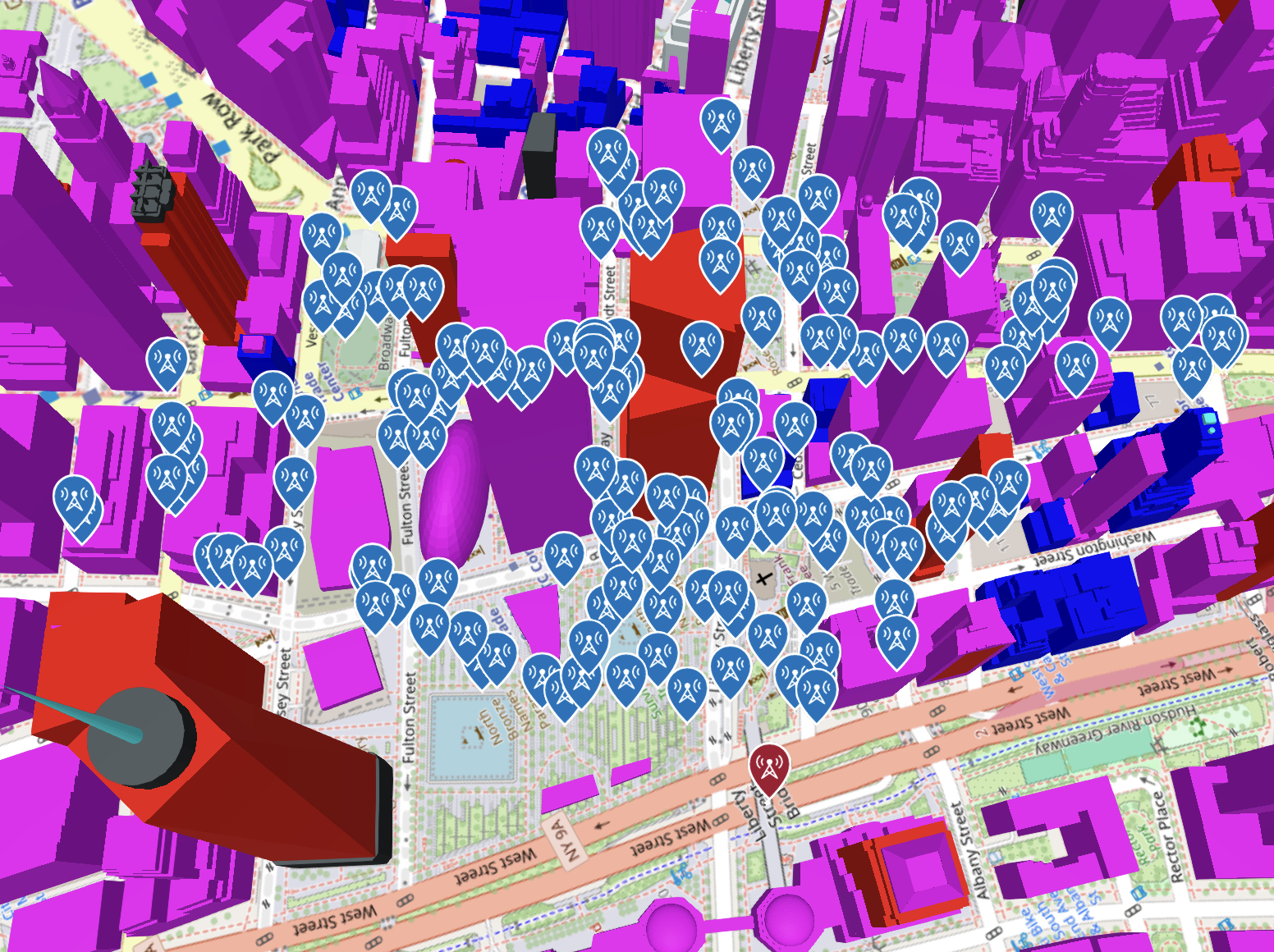}
  }
  \hfill
  \subfloat[Chicago simulation scenario \label{F7b}]{%
    \includegraphics[width=0.48\linewidth]{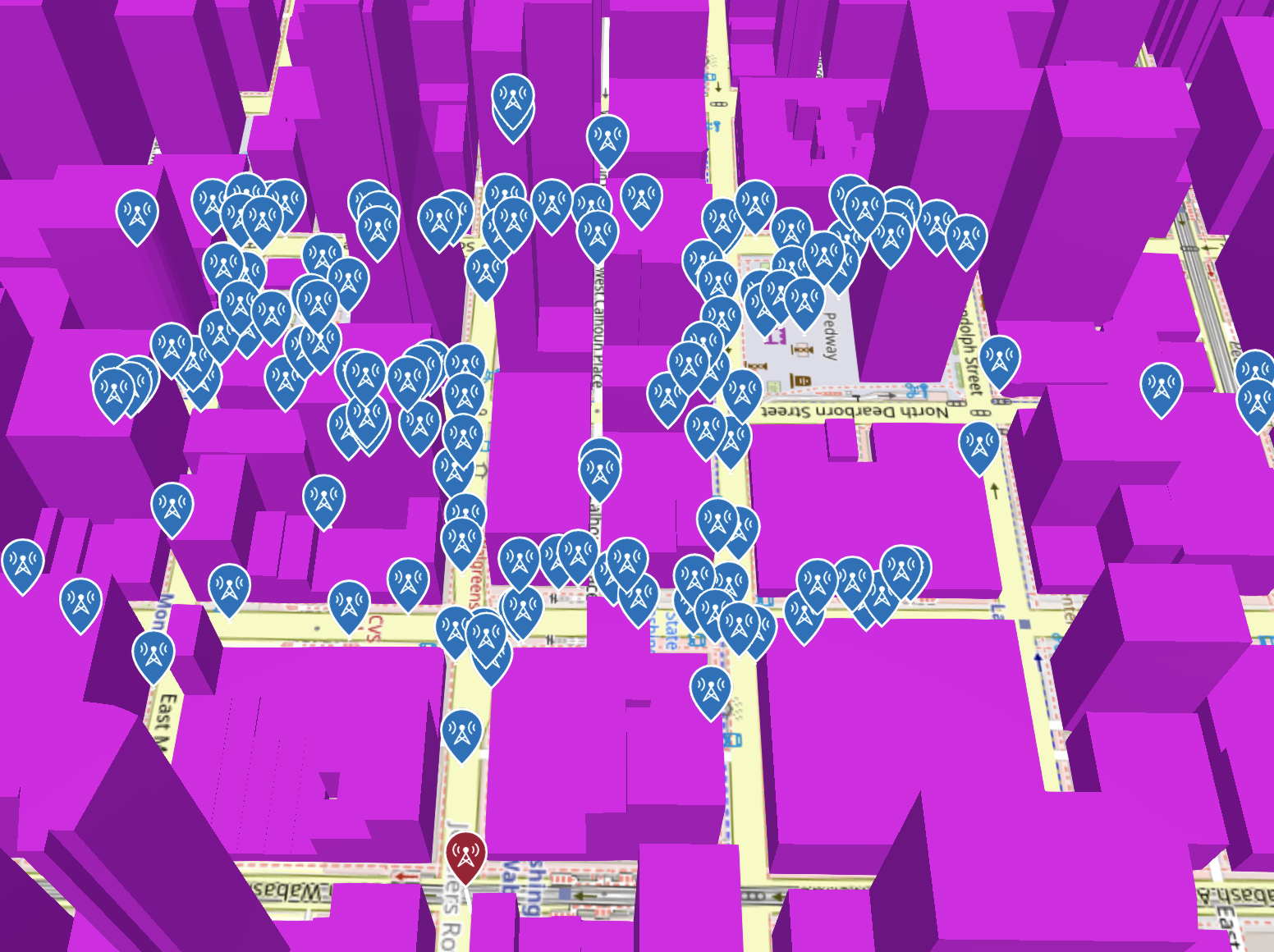}
  }
  
  \caption{Spatial distribution of the gNB (red) and UEs (blue) in both scenarios. All UEs are positioned at a standard height of 1.5 m above ground. 
  Although some blue markers may appear to float above buildings, this is a visualization artifact to ensure visibility of UEs positioned behind tall structures. \looseness=-1
  }
  \label{F1}
  \vspace{-1.5em}
\end{figure}

We use MATLAB's ray tracing (RT) engine to shoot roughly $655K$ rays into the environment. We utilize two distinct urban canyon environments (Manhattan and Chicago), with structural and material properties sourced from OpenStreetMap. Candidate UE locations are generated randomly within a 50 to 500 m radial distance from gNB and a $120^\circ$ azimuth sector relative to the gNB broadside. If a generated coordinate falls inside an existing building footprint, it is discarded; otherwise, a UE is placed at that location at a standard pedestrian height of 1.5m. While this UE placement logic remains identical across both cities, the gNB positioning is intentionally varied to create contrasting propagation conditions. In the Manhattan scenario, the gNB is placed at an open intersection, providing a large area for UE placement in front of the gNB with a strong line-of-sight (LOS) path. Conversely, the Chicago gNB is positioned deep within a narrow street canyon, where users are either restricted to a LOS corridor or heavily shadowed by surrounding high-rises. This environmental disparity is chosen deliberately: it allows us to evaluate the precoding methods across a setup with a prevalence of strong channels (Manhattan) versus a setup with a prevalence of weak, heavily shadowed channels (Chicago). Finally, an initial pool of 200 UEs is placed in both cases and following the RT execution, only UEs that successfully receive at least one valid propagation path from the gNB within three reflections or one diffraction are retained for link level evaluation (Manhattan: 157, Chicago: 135). Fig. \ref{F1} illustrates the gNB and UE placements in both scenarios. \looseness=-1

\subsection{Link Budget Parameters} \label{LINK_BUDGET}

\begin{table}[t]
  \caption{Key Simulation Parameters}
  \label{tab:sim_params}
  \centering
  \footnotesize
  \renewcommand{\arraystretch}{1.1}
  \begin{tabular}{@{}p{0.40\columnwidth}p{0.55\columnwidth}@{}}
    \toprule
    \textbf{Parameter} & \textbf{Value} \\
    \midrule
    \multicolumn{2}{@{}l}{\textit{General}} \\
    Carrier frequency & 3.75~GHz \\
    Bandwidth / SCS & 100~MHz (272 PRBs) / 30~kHz \\
    Transmission scheme & SU-MIMO, full-band allocation per UE \\
    Precoding granularity & Subband, 16 PRBs \\
    \midrule
    \multicolumn{2}{@{}l}{\textit{gNB}} \\
    Antenna array & $\pm45^\circ$ dual-pol. UPA, $8\times12\times2=192$ elements, $0.5\lambda$ spacing~\cite{ITU-R_5D989_Annex4.4} \\
    Virtualization & $1\times6$ subarrays $\rightarrow$ 32 ports ($N_1=8$, $N_2=2$) \\
    Port spacing (H / V) & $0.5\lambda$ / $3\lambda$ \\
    Element gain / pattern & 6.4~dBi incl.\ ohmic loss~\cite{3GPP_TR38922} / 3GPP TR 38.901 standard \cite{3GPP_TR38901} \\
    Height / downtilt & 25~m / $6^\circ$ (mechanical) \\
    Conducted power & 38~dBm/port; 53~dBm total per 100~MHz \\
    EIRP & $\approx$79~dBm per 100~MHz \\
    CSI-RS configuration & Row = 16, Density = 1, Boost = 6 dB \\
    Noise Figure & 5~dB \\
    Implementation loss (UL) & 5~dB \\
    \midrule
    \multicolumn{2}{@{}l}{\textit{UE}} \\
    Antenna configuration & $\pm45^\circ$ dual-pol.\ pairs, 150~mm apart, 4 ports~\cite{11389146} \\
    Element gain / HPBW / SLA & 2.3~dBi incl.\ ohmic loss / $125^\circ$ / 22.5~dB \\
    Height / orientation & 1.5~m / random yaw, zero pitch \\
    Transmit power & 23~dBm (Power Class 3) \\
    SRS configuration & $K_{\text{TC}}=2 , C_{\text{SRS}}=63 , B_{\text{SRS}}=B_{\text{hop}=0} $ \\
    Noise Figure & 7~dB \\
    Implementation loss (DL) & 6~dB \\
    \midrule
    \multicolumn{2}{@{}l}{\textit{Ray tracing}} \\
    Engine / rays & MATLAB ray tracer / $\approx$655k rays \\
    Max.\ interactions & 3 reflections or 1 diffraction \\
    Max absolute path loss & 150 dB \\
    Max relative path loss & 25 dB \\
    \bottomrule
  \end{tabular}
\end{table}

At the gNB, each of the 32 logical ports is driven by a power amplifier with an assumed limit of 38 dBm per 100 MHz, yielding a total gNB conducted power of 53 dBm per 100 MHz. Given the maximum element gain of 6.4 dBi, the effective isotropic radiated power (EIRP) for the gNB is $53+6.4+10\log_{10}(96)\approx79$ dBm per 100 MHz. The FCC restricts EIRP to 62 dBm/MHz (equivalent to 82 dBm per 100 MHz) \cite{FCC_LowerCband_2020}. While this allows the conducted power to reach 56 dBm for broader coverage, we restrict it to 53 dBm —a value typical of practical macro-cell deployments  that adequately serves our 500 m simulation radius \cite{ITU-R_5D_Saudi_7GHz_2025 , elnashar2026bridging , feltrin2021potential}. For UL reception, the gNB noise figure (NF) is set to 5 dB \cite{3GPP_TS38104}. 

For the UE, we assume a total conducted power of 23 dBm, corresponding to the standard 3GPP Power Class 3 limit. With an element gain of 2.3 dBi after ohmic loss, the resulting UE EIRP is 25.3 dBm, which operates safely below the FCC's 30 dBm limit \cite{FCC_LowerCband_2020}. The UE NF for DL reception is set to 7 dB \cite{ITU-R_IMT2030_EVAL}. In addition, for DL and UL transmissions, implementation losses of 6 dB and 5 dB are applied, respectively. \looseness=-1 This 1 dB higher penalty in the DL primarily accounts for the stricter hardware constraints and form-factor limitations of UE receivers, alongside a higher susceptibility to interference. The key simulation parameters are summarized in Table~\ref{tab:sim_params}. \looseness=-1 

It is important for subsequent discussion to consider the symbol power difference between the gNB-transmitted CSI-RS and UE-transmitted SRS. Given the 38 dBm per-port conducted power over a 100 MHz bandwidth (272 PRBs), the baseline gNB power per subcarrier is $38 - 10\log_{10}(272\times12)\approx2.86$ dBm. 
With a typical 6 dB boost, the effective CSI-RS symbol power reaches 8.86 dBm. In contrast, for UL SRS transmission the UE's 23 dBm total transmit power limit is distributed across its four antenna ports for SRS transmission, resulting in a per-port limit of 17 dBm. Assuming the SRS is transmitted across the full 100 MHz channel with a comb parameter of $K_{\text{TC}}=2$ (i.e., mapped to every second subcarrier), the resulting SRS symbol power drops to $17-10\log_{10}(\frac{272\times12}{2})\approx-15.13$ dBm. Consequently, the CSI-RS symbols are transmitted with roughly a 24 dB higher power than the SRS symbols.

\section{Link Level Simulation Framework}
We consider a single-user MIMO (SU-MIMO) setup where the full 100 MHz channel bandwidth at 3.75 GHz center frequency is allocated to each user. 

To evaluate DL performance, we consider the two distinct 3GPP-compliant transmission schemes introduced in Section~\ref{sec:introduction}.
For CSIRS-PMI, this work considers Type-I single-panel codebook, which is the most widely deployed in commercial networks. The UE selects the optimal precoding matrix from the codebook based on its channel estimate and feeds the PMI back to the gNB. Although PMI selection process at UE is left to implementation~\cite{3GPP_TS38214}, in our simulation the UE reports the precoder index that maximizes the post-equalization SINR similar to \cite{matlab_tutorial}. Additionally, we consider a 6 dB CSI-RS power boost.
For SRS-SVD, the UE transmits full-band SRS without frequency hopping and the UL channel is estimated at the gNB. Assuming perfect hardware calibration and channel reciprocity, the gNB estimates the DL channel, performs SVD, and constructs the DL precoder using the dominant right-singular vectors. Because commercial estimators are proprietary, we process both CSI-RS and SRS based channel estimation using MATLAB's \texttt{nrChannelEstimate} function \cite{matlab_tutorial}, applying least squares (LS) estimation at pilot locations followed by smoothing and time-frequency interpolation. Both schemes use subband precoding with a 16 PRB granularity. For subband SVD, channel matrices are averaged across subcarriers within each subband before applying SVD similar to \cite{matlab_tutorial2}.
Finally, we sweep the transmission layers from 1 to 4, extracting the best Type-I PMI or SVD-based precoder for each respective rank. \looseness=-1

\begin{figure*}[t]
  \centering
  \includegraphics[width=\textwidth]{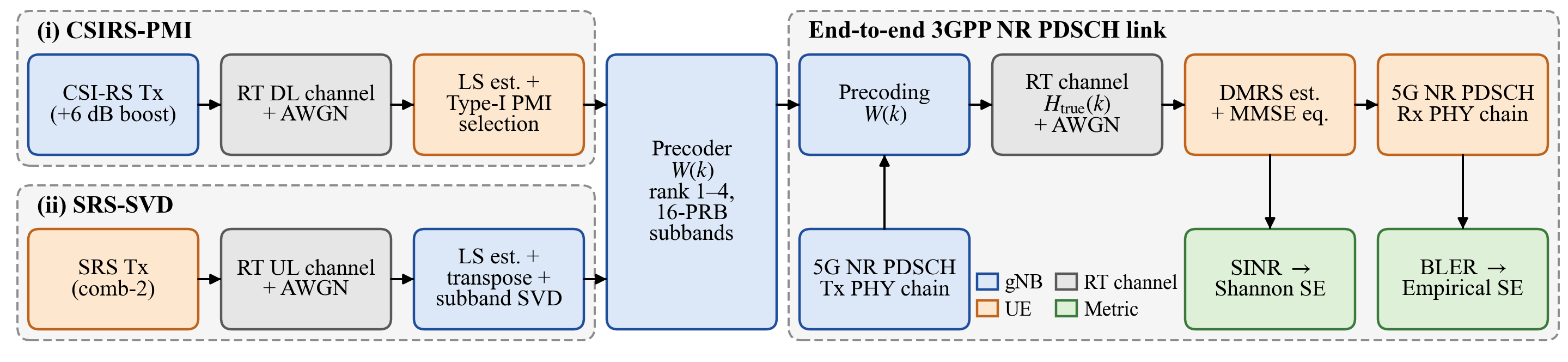}
  \caption{Link-level simulation framework: CSI acquisition via (i) CSI-RS with Type-I PMI feedback or (ii) SRS with SVD precoding, followed by an end-to-end 3GPP NR PDSCH chain over the ray-traced channel.}
  \label{fig:lls_block}
  \vspace{-1em}
\end{figure*}

In the simulation setup, for a given layer, modulation and coding scheme (MCS), Transport Block size is first calculated and random bits of payload are generated accordingly. This payload is processed through the complete 5G NR physical layer transmit chain—encompassing CRC (cyclic redundancy check) attachment, code block (CB) segmentation, LDPC coding, rate matching, interleaving, CB concatenation, scrambling, symbol modulation, and layer mapping—and is subsequently multiplexed with DMRS symbols to form a transmission grid of dimensions $N_{\text{sc}} \times N_{\text{symbol}} \times N_{\text{layers}}$, corresponding to the allocated subcarriers, a standard 14-symbol slot, and the spatial layers, respectively. 

For a given subcarrier k, let the true channel be denoted by $H_{\text{true}}(k)$ and the estimated channel—derived from either the CSI-RS or SRS—by $H_{\text{est}}(k)$. Let W(k) be the precoding matrix constructed from $H_{\text{est}}(k)$ using either the PMI or SVD. Then the received DL signal y(k) at subcarrier k can be expressed as: $y(k) = H_{\text{true}}(k)W(k)x(k)+n(k)$ where $n(k)$ represents the additive white Gaussian noise (AWGN) with a variance $N_0$ ($N_0$ calculated based on the UE NF and the DL implementation margin) and $x(k)$ represents the transmitted DL data at subcarrier k. \looseness=-1

Upon reception of the DL slot, the UE estimates the effective DL channel (incorporating the gNB precoder) and noise variance from known DMRS. If $\hat{H}_{\text{est}}(k)$ is the UE estimate of effective channel at subcarrier k (true effective channel $\hat{H}_{\text{true}}(k)=H_{\text{true}}(k)W(k)$) and $N_{0,\text{est}}$ is the noise variance estimate, then the Minimum Mean Square Error (MMSE) filter is constructed as $G_{\text{MMSE}}(k) =(\hat{H}_{\text{est}}(k)\hat{H}_{\text{est}}^H(k)+N_{0,\text{est}}I)^{-1}\hat{H}_{\text{est}}^H(k)$ which is utilized to extract the equalized data symbols. Subsequently, the equalized symbols are passed through the complete PDSCH receive chain, sequentially reversing the transmitter operations (layer demapping, demodulation, rate recovery, and LDPC decoding).
Finally, following CRC validations at both the code block and transport block levels, the transport block is ultimately declared as either successfully received or in error based on whether CRC check has passed or failed, providing the basis for the system's block error rate (BLER) evaluation.
As performance metrics we log two forms of SE: \looseness=-1

\textbf{i) Empirical SE}: To determine the empirical 3GPP-specific achievable SE for a given UE $u$, we evaluate the performance across all candidate spatial layers $l \in \{1, 2, 3, 4\}$. For each layer candidate $l$, we sweep through the MCS indices ranging from 0 to 27 from 3GPP MCS Table 2 (256 QAM table) \cite{3GPP_TS38214}. The BLER for each index is evaluated via Monte Carlo simulations with minimum 100 block error collection criterion. Through this process, we identify the highest sustainable MCS index, denoted as $\beta$. This index is selected to satisfy the standard $10\%$ BLER threshold, such that the BLER at MCS $\beta$ is less than 10\%, while increasing to $\beta + 1$ results in a BLER exceeding 10\%. The achievable SE for this specific layer configuration is then computed as $SE_l = l \times SE(\beta)$, where $SE(\beta)$ is the SE corresponding to MCS $\beta$. Finally, the optimal spatial rank $R_u$ and the highest achievable SE $SE_u$ for the user are calculated by maximizing across all layer configurations:
\begin{equation*}
    SE_u = \max_{l \in \{1,2,3,4\}} SE_l,
    R_u = \text{argmax}_{l \in \{1,2,3,4\}} SE_l
\end{equation*}

\textbf{ii) Shannon SE}: 
In addition to the empirical SE, we also look at the Shannon SE for a given applied precoder taking the imperfect channel knowledge at UE into consideration. For notational brevity, the subcarrier index $(k)$ is intentionally omitted in the following discussion. Applying the calculated MMSE filter $G_{\text{MMSE}}$ (based on estimated channel) to the received signal $y$ yields the equalized symbol $\hat{x}$:
$\hat{x} = G_{\text{MMSE}}~y = M x + G_{\text{MMSE}} n$
where $M = G_{\text{MMSE}} \hat{H}_{true}$ represents the combined channel and MMSE equalizers response matrix. Extracting the signal for the $p$-th spatial layer, we get the desired signal component, the inter-layer interference, and the post-equalization noise:
\begin{equation*}
    \hat{x}(p) = M(p,p)x(p) + \sum_{q \neq p} M(p,q)x(q) + G_{\text{MMSE}}(p,:)n
\end{equation*}
Consequently, the post-equalization SINR for layer $p$ can be calculated as:
\begin{equation*}
    \text{SINR}(p) = \frac{|M(p,p)|^2}{\sum_{q \neq p} |M(p,q)|^2 + \|G_{\text{MMSE}}(p,:)\|^2 N_0}
\end{equation*}
The Shannon SE for this specific subcarrier is then calculated by summing the Shannon capacities across all layers:
$SE = \sum_{p=1}^{N_{\text{layers}}} \log_2(1 + \text{SINR}(p))$.
Finally, averaging these subcarrier-specific SE values across the bandwidth yields the wideband SE. As with the empirical SE case, this wideband capacity is evaluated across all candidate spatial ranks, with the maximum value selected as the user's final Shannon SE. The link-level simulation framework is illustrated in Fig. \ref{fig:lls_block}.

\color{black}
\section{Results}
\subsection{SE Comparison}

\begin{figure}[t]
  \centering
  \subfloat[Manhattan simulation scenario \label{F7a}]{%
    \includegraphics[ height = 4.5cm ]{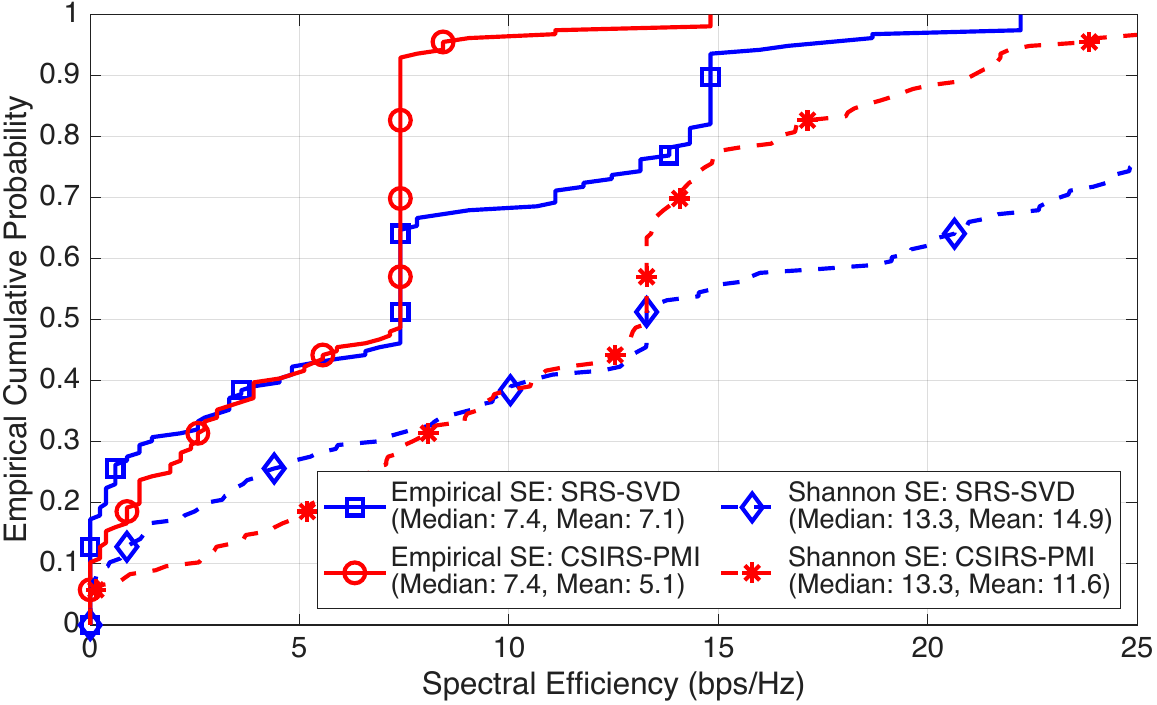}
  }
  \\
  \subfloat[Chicago simulation scenario \label{F7b}]{%
    \includegraphics[height = 4.5cm ]{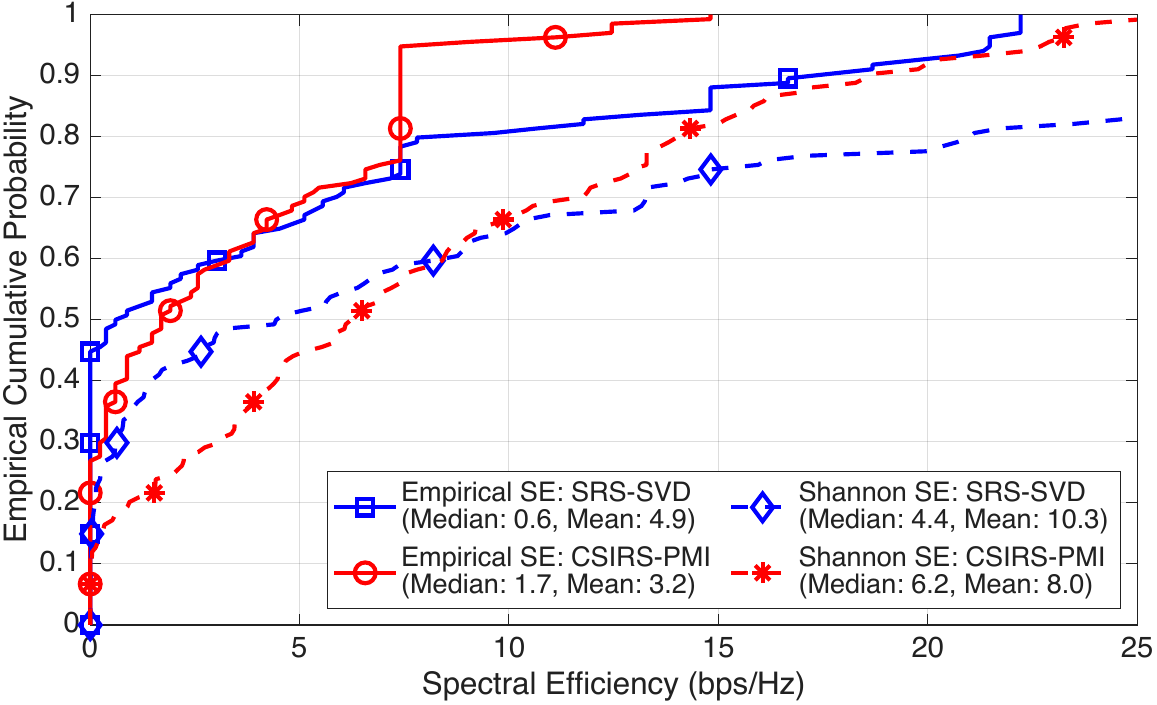}
  }
  
  \caption{Comparison of empirical and Shannon SE between SRS-SVD and CSIRS-PMI on (a) Manhattan and (b) Chicago deployment. It can be observed that SRS-SVD provides superior performance in higher SE regime whereas CSIRS-PMI becomes preferable for users in the low-SE regime. 
  \looseness=-1 }
  \label{F4}
  \vspace{-1.5em}
\end{figure}

First, we compare the SE of the two DL transmission schemes across both deployments. As can be seen in Fig. \ref{F4}, in both deployments, there is no single transmission scheme that performs uniformly better. In the low-SE regime, CSIRS-PMI yields higher capacity, whereas this trend reverses in the high-SE regime, where SRS-SVD dominates.

This performance crossover is fundamentally driven by reference symbol transmit power asymmetry and the resulting channel estimation accuracy gap. While SVD offers superior beamforming over the quantized Type-I PMI codebook, it relies on UL SRS. Due to tight UE power constraints, SRS-based estimates suffer larger errors than high-power CSI-RS. For edge users, this estimation bottleneck nullifies SVD's beamforming gain, making CSIRS-PMI optimal. Conversely, for users with strong channels, estimation error is no longer a bottleneck, allowing the unquantized SRS-SVD precoder to achieve superior SE. This is validated by the channel estimation Normalized Mean Square Error (NMSE). In Manhattan, CSIRS-PMI optimal users exhibit severely degraded estimation (where noise dominates the signal, yielding a positive median SRS NMSE of 18 dB vs. 3 dB for CSI-RS). In contrast, SRS-SVD optimal users possess highly accurate estimates (median SRS NMSE of -15 dB vs. -29 dB for CSI-RS), allowing SVD beamforming to dictate performance. \looseness=-1

As observed in Fig. \ref{F4}, the Shannon SE is larger in magnitude than the empirical SE which is attributed to three primary factors. First, empirical SE is capped at 7.4 bps/Hz per layer by 3GPP MCS Table 2 (achieved via 256 QAM), whereas unconstrained Shannon capacity can reach 13–14 bps/Hz at high SNRs (30–40 dB). Second, Shannon capacity assumes independent rate adaptation per layer and subcarrier, while 3GPP standards restrict empirical SE to a single shared MCS across the entire transport block. Finally, an inherent gap remains between modern LDPC codes and the theoretical Shannon limit. Crucially, despite this magnitude difference, the relative performance trend is similar in both cases. Both capacity metrics exhibit the exact same crossover behavior and show that the optimal DL transmission method is different in low SE and high SE regime. \looseness=-1

In the Manhattan deployment, which has a prevalence of users with strong channel conditions, a larger number of users (63) achieve better empirical SE with SRS-SVD compared to users where CSIRS-PMI is optimal (45). For the remaining users, the empirical SE is the same under both methods. With SRS-SVD spatial multiplexing is significantly more viable: a transmission rank greater than one (Rank $>$ 1) is optimal for 35\% of users, whereas this ratio falls to 7\% for CSIRS-PMI. SRS-SVD has a larger overall mean SE (7.1 bps/Hz compared to 5.1 bps/Hz for CSIRS-PMI) due to the high SE it achieves for strong users. However, CSIRS-PMI serves edge users better. For the bottom 30\% of users, the mean SE for SRS-SVD is 0.2 bps/Hz, whereas with CSIRS-PMI it is 0.7 bps/Hz, indicating that CSIRS-PMI provides better support for these users. \looseness=-1

By contrast, in the Chicago deployment, we observe a near tie between SRS-SVD preferred users (43) and CSIRS-PMI preferred users (42). With SRS-SVD, Rank $>$ 1 is optimal for 23\% of users, whereas the ratio falls to 6\% for CSIRS-PMI. In terms of outage—defined as users where even the lowest MCS could not be decoded with less than a 10\% BLER—CSIRS-PMI results in a 27\% outage rate, whereas SRS-SVD has a 45\% outage rate. This shows that CSIRS-PMI can be the preferred method to reduce outages in deployments where there is a possibility of very weak channels.

\begin{figure*}[t]
  \centering
  \subfloat[Manhattan simulation scenario \label{F7a}]{%
    \includegraphics[width=0.45\linewidth]{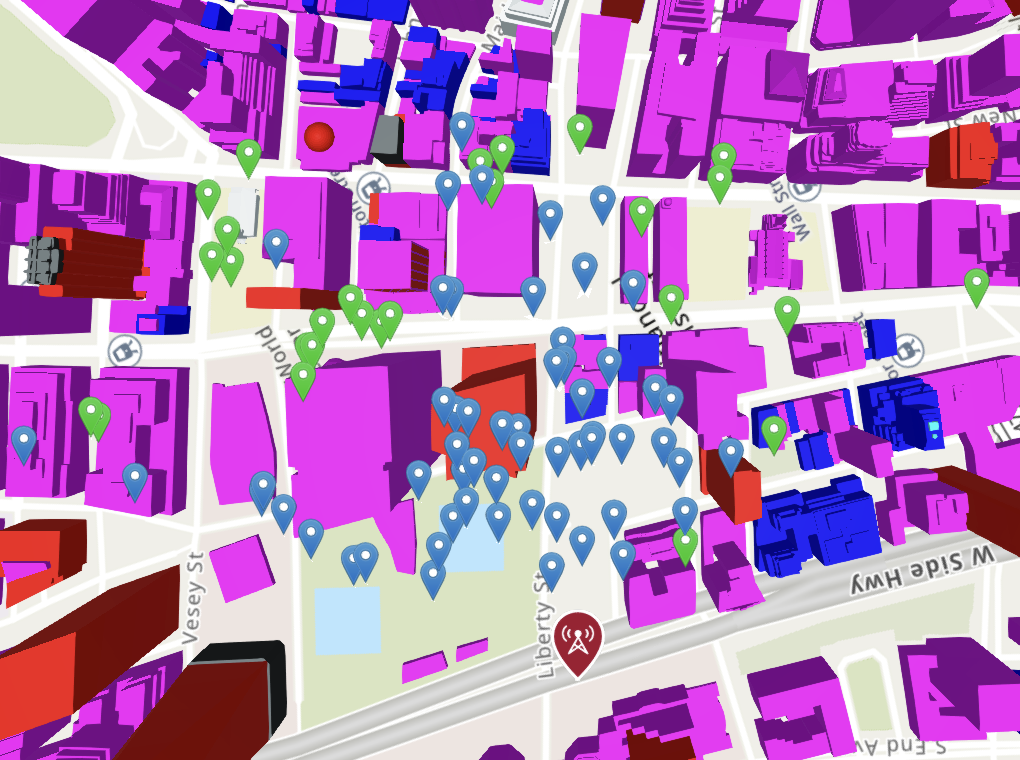}
  }
  \hfill
  \subfloat[Chicago simulation scenario \label{F7b}]{%
    \includegraphics[width=0.45\linewidth]{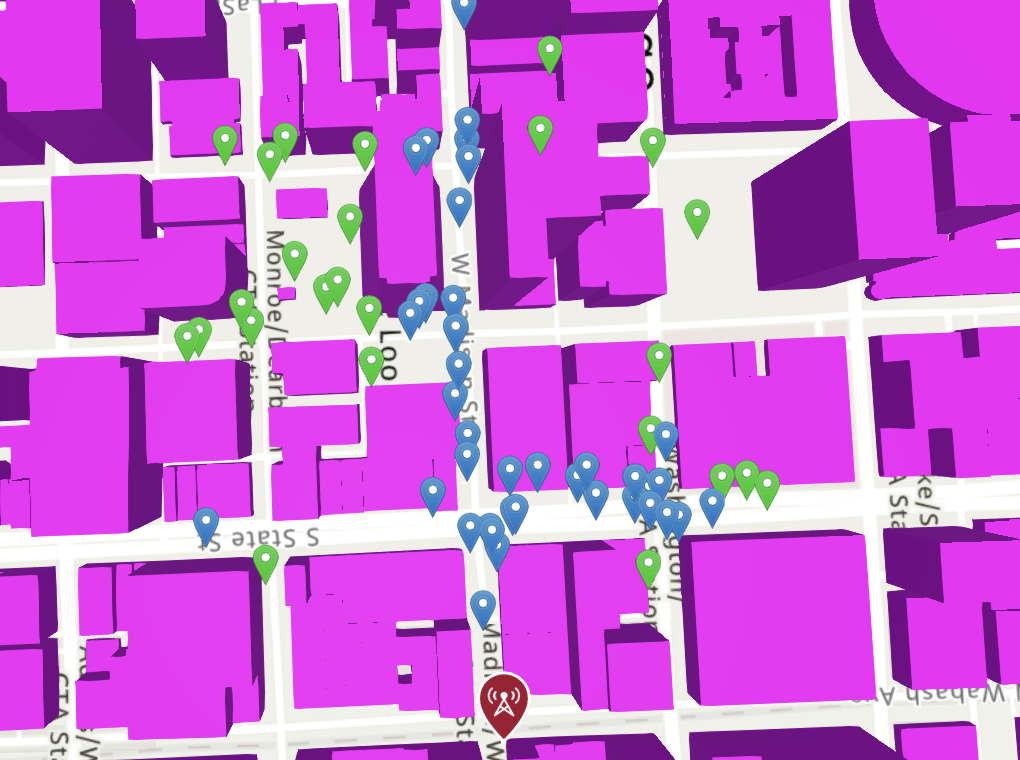}
  }

  \caption{ Spatial distribution of UEs color-coded by their dominant precoding scheme on (a) Manhattan and (b) Chicago deployment. Blue and green markers denote locations where SRS-SVD and CSIRS-PMI, respectively, outperform the alternative scheme by a margin of at least 0.5 bps/Hz in terms of empirical SE. \looseness=-1 }
  \label{F5}
  \vspace{-1em}
\end{figure*}

Fig. \ref{F5} displays the spatial distribution of users where one transmission scheme outperforms the other by 0.5 bps/Hz or more (blue markers: SRS-SVD optimal, green markers: CSIRS-PMI optimal). In the Manhattan deployment, most of the user locations where SRS-SVD outperforms are in close proximity to the gNB. Similarly, in Chicago, the user locations where SRS-SVD outperforms are either radially close to the gNB or in a straight corridor in front of it, benefiting from a strong LOS component. In both cases, the user locations where CSIRS-PMI outperforms are radially distant or heavily shadowed behind structures. \looseness=-1

\subsection{Feature-Based Rule}
We investigate whether the gNB can determine the optimal transmission scheme based on standard metrics reported by the UE, such as CSI-RSRP and CQI. CSI-RSRP is calculated as the average received signal power of the CSI-RS across all reference symbol locations, while CQI is a 16-level discrete feedback metric computed by the UE for link adaptation. In our simulation, we model the CQI calculation as follows: the UE first computes the effective channel, assuming the optimal PMI matrix and rank will be applied at the gNB. The UE then calculates the post-MMSE equalized subcarrier SINRs, which are compressed into a single effective scalar using Exponential Effective SINR Mapping (EESM). This EESM SINR is finally mapped to the highest CQI index satisfying a 10\% BLER target. Although exact CQI computation algorithms in commercial UEs are proprietary, this EESM-based methodology is widely adhered to in academic literature and standard-compliant simulations \cite{vienna, matlab_tutorial}.
The inductive bias behind investigating these features is that they directly reflect channel strength and quality, which as discussed previously, are the primary factors dictating the optimal transmission scheme. In addition, while not strictly mandatory for every 5G operation, these parameters are routinely available at the gNB.

\begin{figure}[t]
    \centering
    \includegraphics[ width=1\linewidth ]{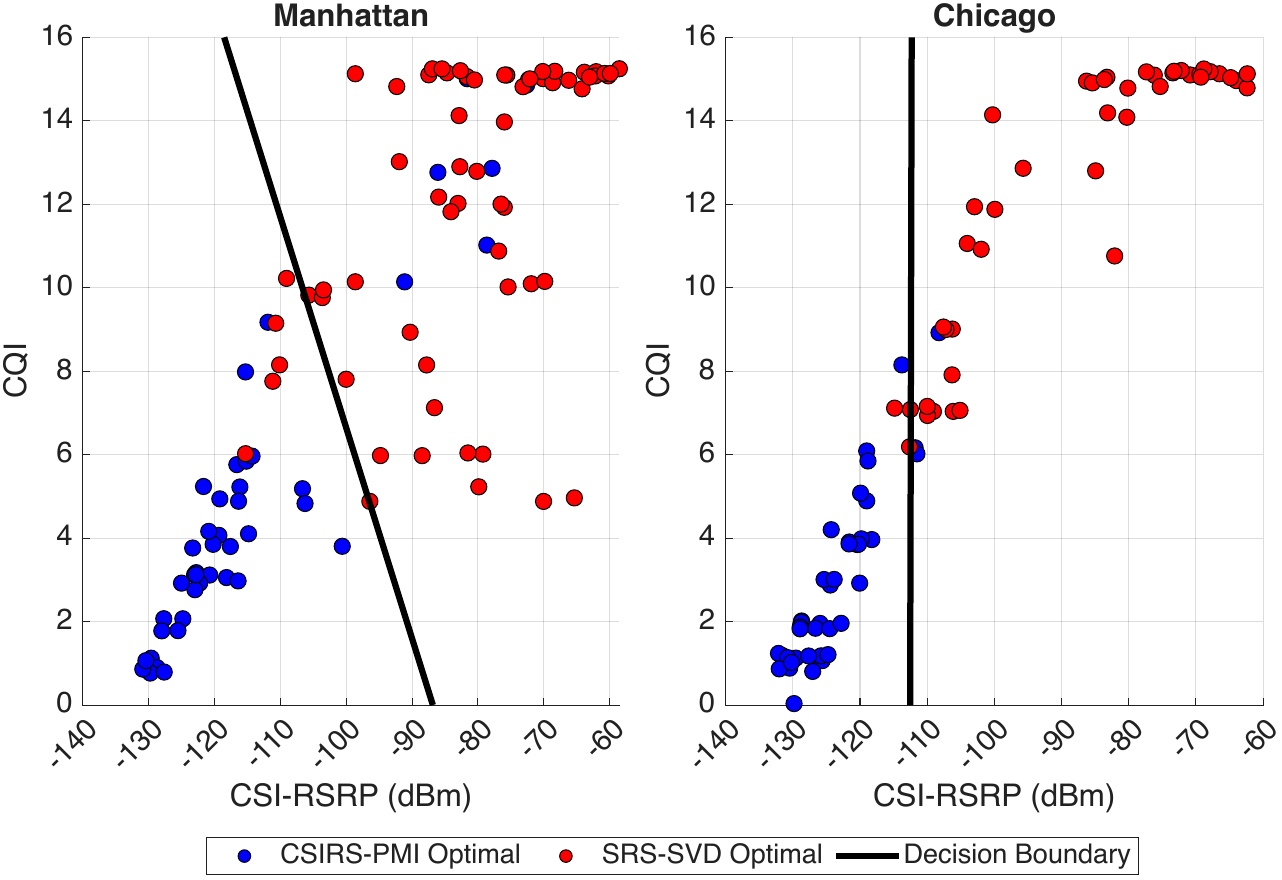}
    \caption{Scatter plot of UEs categorized by their optimal precoding scheme with respect to CSI-RSRP and CQI in Manhattan and Chicago. The solid black line represents the derived logistic regression decision boundary separating the CSIRS-PMI and SRS-SVD optimal regions. 
    A small jitter is added to CQI values to prevent points from overlapping. 
    \looseness=-1
    }
    \vspace{-1em}
    \label{F6}
\end{figure}

\begin{figure}[t]
    \centering
    \includegraphics[height = 4.5cm ]{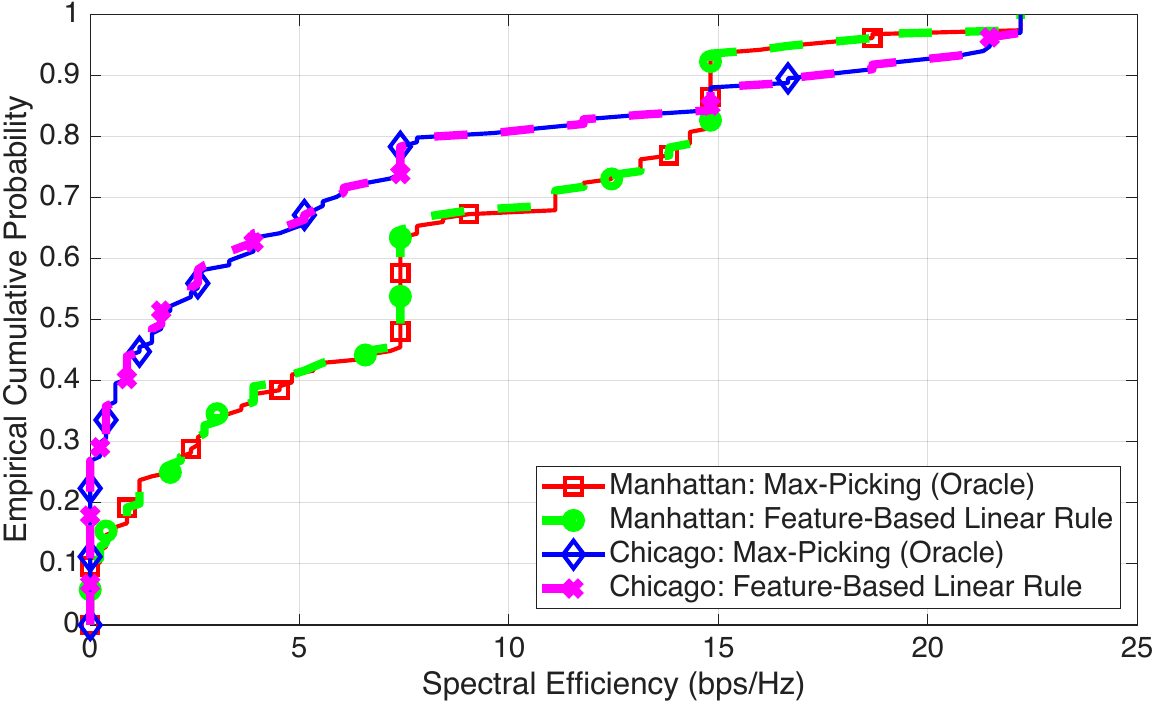}
    \caption{Comparison of empirical SE achieved by the Max-Picking oracle and feature (CSI-RSRP and CQI) based linear classifier. The near-perfect overlap of the curves indicates that the linear classifier is successfully able to choose the capacity-maximizing precoding scheme in both simulated environments.
    }
    \vspace{-1.5em}
    \label{F7}
\end{figure}

Upon investigation, we observe that across both deployments, CSI-RSRP and CQI provide a linear boundary that effectively separates the optimal transmission schemes, as shown in Fig. \ref{F6}. Interestingly, a simple horizontal or vertical threshold on a single feature provides almost equally good separation.
Using these two metrics as features and applying 5-fold cross-validation on the given data, we achieved around 90\% accuracy (Manhattan: 90\%, 95\% bootstrap confidence interval (CI): 83--94\%; Chicago: 93\%, 95\% CI: 87--98\%; excluding tied users to prevent inflating accuracy) in predicting the optimal precoding scheme for a UE using simple logistic regression. Notably, a static threshold on individual features (e.g., choose CSIRS-PMI if CSI-RSRP $\le -112$~dBm) also achieves close to 90\% accuracy.
In Fig. \ref{F7}, we compare the oracle max-picking SE (where the highest SE between SRS-SVD and CSIRS-PMI is chosen for each UE) with the SE achieved using the logistic regression predictions. Across both deployments, we observe a close overlap (suggesting that misclassified UEs have relatively similar performance under either method), demonstrating that with a simple, low-compute linear prediction rule, the gNB can maintain high SE for users with strong channels while also effectively serving edge users.\looseness=-1

While CSI-RSRP and CQI require UE feedback, SRS-RSRP is directly available at the gNB (provided SRS is enabled). In our experiments, similar prediction performance is achieved using SRS-RSRP in place of CSI-RSRP, which is expected since both metrics depend on overall channel strength. This flexibility allows the network to optimize its default transmission strategy based on the deployment scenario. For example, if a gNB covers a small radius where UEs are mostly close by (e.g. small cell/urban micro), it can set SRS as its default reference signal. The gNB can then monitor the available SRS-RSRP and decide to switch to CSIRS-PMI only for users where the prediction rule suggests it is necessary. By contrast, a gNB covering a broader area with many poor-channel UEs (e.g., rural macro) can default to CSI-RS. In this case, the gNB relies on UE CSI-RSRP feedback, switching to SRS-SVD only when the channel is strong enough to benefit. \looseness=-1

\subsection{SNR vs BLER on Statistical Channel}
In addition to the RT environments, we also evaluate performance over normalized statistical channels. Specifically, we utilize the CDL-C channel, which models non-line-of-sight (NLOS) environments with multipath scattering. The delay spread is set to 300 ns, and the remaining system parameters are kept consistent with our previous simulations. \looseness=-1 

The channel estimation SNR differs from the PDSCH data SNR, since reference signals and data are not transmitted at the same power.
In the CSIRS-PMI scheme, the CSI-RS is boosted by 6~dB relative to the PDSCH data symbols. By contrast, in the SRS-SVD scheme the SRS is transmitted by the UE, whose power is far lower than that of the gNB. As shown in Section~\ref{LINK_BUDGET}, the SRS arrives with 24~dB less power per subcarrier than the CSI-RS.
Because the SRS is received at the gNB, which has a lower NF and implementation loss than the UE, this gap shrinks to an effective SNR difference of about 21~dB. In our simulation,
we therefore set the CSI-RS and SRS channel estimation SNRs to $x+6$~dB and $x-15$~dB, respectively, for a PDSCH data SNR of $x$~dB.

\begin{figure}[t]
    \centering
    \includegraphics[height = 4.5cm ]{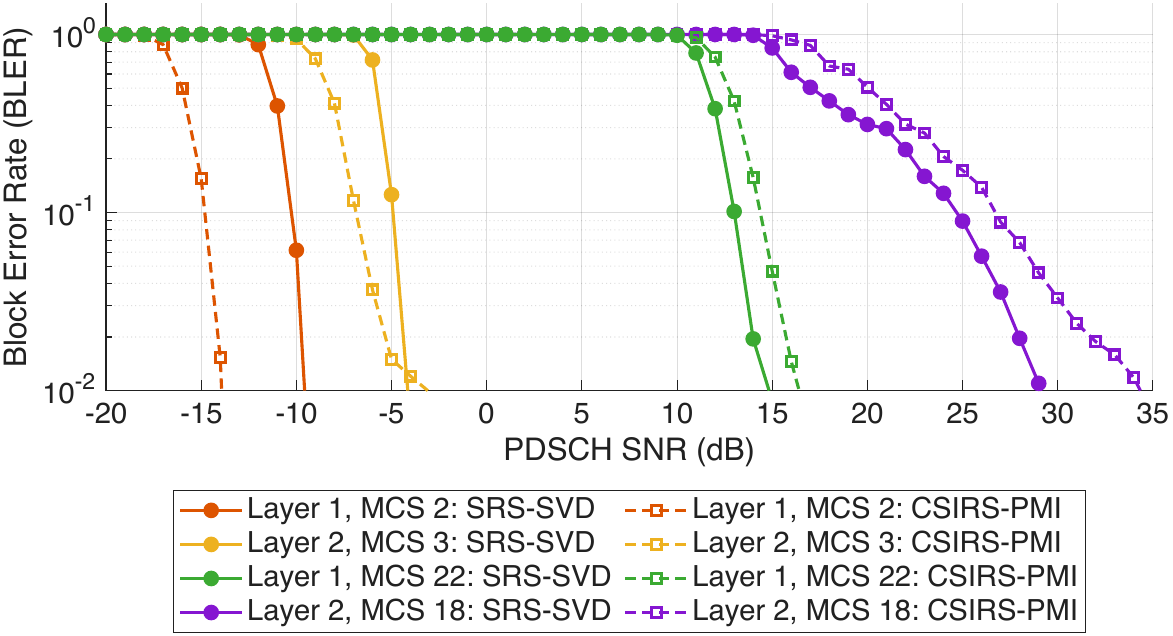}
    \caption{SNR vs BLER Comparison on CDL-C channel for SRS-SVD and CSIRS-PMI considering the difference in channel estimation SNR from PDSCH SNR explicitly. 
    \looseness=-1}
    \vspace{-1.5em}
    \label{F8}
\end{figure}

Fig.~\ref{F8} shows the BLER for different layer and MCS configurations.
At low MCS, which operates at low SNR, CSIRS-PMI reaches the BLER waterfall earlier, as the inaccurate SRS-based estimate offsets the SVD beamforming gain.
Conversely, at high MCS (high SNR regime) the SRS estimate is accurate enough to exploit the unquantized SVD precoder, and SRS-SVD performs better.
This holds for both one and two layers, confirming the ray-tracing observation that the preferred scheme is determined by the operating SNR, and thus by the channel conditions of each user. \looseness=-1

\section{Conclusion}

This paper compares SRS-based SVD precoding and CSI-RS-based Type-I PMI precoding for mid-band MIMO through an end-to-end, 3GPP-compliant link-level simulation over RT channels in Manhattan and Chicago. The results show that neither scheme is uniformly superior. Because SRS is transmitted at the limited UE power, roughly 24~dB below CSI-RS per symbol, SRS-based channel estimates degrade sharply for weak users. As a result, CSIRS-PMI provides higher SE for edge users (0.7 vs.\ 0.2~bps/Hz for the bottom 30\% in Manhattan) and fewer outages (27\% vs.\ 45\% in Chicago). For users with strong channels, accurate SRS estimates allow SVD precoding to outperform the quantized codebook, raising the mean SE in Manhattan from 5.1 to 7.1~bps/Hz and making multi-layer transmission optimal for 35\% of users, compared with 7\% for CSIRS-PMI. The same crossover appears on the CDL-C statistical channel. We further showed that a low complexity logistic regression on standard CSI-RSRP and CQI reports, or on SRS-RSRP available at the gNB, selects the better scheme with about 90\% accuracy and achieves SE close to the max-picking upper bound. 
Our future work will extend the analysis to different panel sizes and frequencies, while also relaxing the baseline assumptions of this study to evaluate the impact of SRS hopping, higher-resolution Type-II codebooks, frequency scheduling, and non-ideal TDD reciprocity. \looseness = -1

\textbf{AI Usage Disclosure}: Generative AI tools were utilized during the preparation of this paper for polishing sentence structure and improving grammatical flow. 

\vspace{-0.3em}
\bibliographystyle{IEEEtran}
\bibliography{references}

\end{document}